\documentclass[twoside,11pt]{article}

\usepackage{blindtext}

\usepackage{jmlr2e}
\usepackage{amsmath}

\ShortHeadings{Asymmetric Transformer Decoder}{B. Brattoli, J. Shi et al.}
\firstpageno{1}

\begin{document}

\title{Identifying actionable driver mutations in lung cancer using an efficient Asymmetric Transformer Decoder}

\author{\name Biagio Brattoli* \email biagio@lunit.io
\AND
    \name Jack Shi* \email jack@lunit.io
\AND
    \name Jongchan Park \email jcpark@lunit.io 
\AND
    \name Taebum Lee \email taebum.lee@lunit.io 
\AND
    \name Donggeun Yoo \email dgyoo@lunit.io 
\AND
    \name Sergio Pereira \email sergio@lunit.io \\ \\
    \addr Lunit Inc, Seoul, South Korea}

\editor{My editor}

\maketitle

\begin{abstract}
Identifying actionable driver mutations in non-small cell lung cancer (NSCLC) can impact treatment decisions and significantly improve patient outcomes. Despite guideline recommendations, broader adoption of genetic testing remains challenging due to limited availability and lengthy turnaround times. Machine Learning (ML) methods for Computational Pathology (CPath) offer a potential solution; however, research often focuses on only one or two common mutations, limiting the clinical value of these tools and the pool of patients who can benefit from them. This study evaluates various Multiple Instance Learning (MIL) techniques to detect six key actionable NSCLC driver mutations: ALK, BRAF, EGFR, ERBB2, KRAS, and MET ex14. 
Additionally, we introduce an Asymmetric Transformer Decoder model that employs queries and key-values of varying dimensions to maintain a low query dimensionality. This approach efficiently extracts information from patch embeddings and minimizes overfitting risks, proving highly adaptable to the MIL setting. 
Moreover, we present a method to directly utilize tissue type in the model, addressing a typical MIL limitation where either all regions or only some specific regions are analyzed, neglecting biological relevance.
Our method outperforms top MIL models by an average of $3\%$, and over $4\%$ when predicting rare mutations such as ERBB2 and BRAF, moving ML-based tests closer to being practical alternatives to standard genetic testing.
\end{abstract}


\section{Introduction}

\begin{figure*}[t!]
    \centering
    \includegraphics[width=1.0\textwidth,,keepaspectratio]{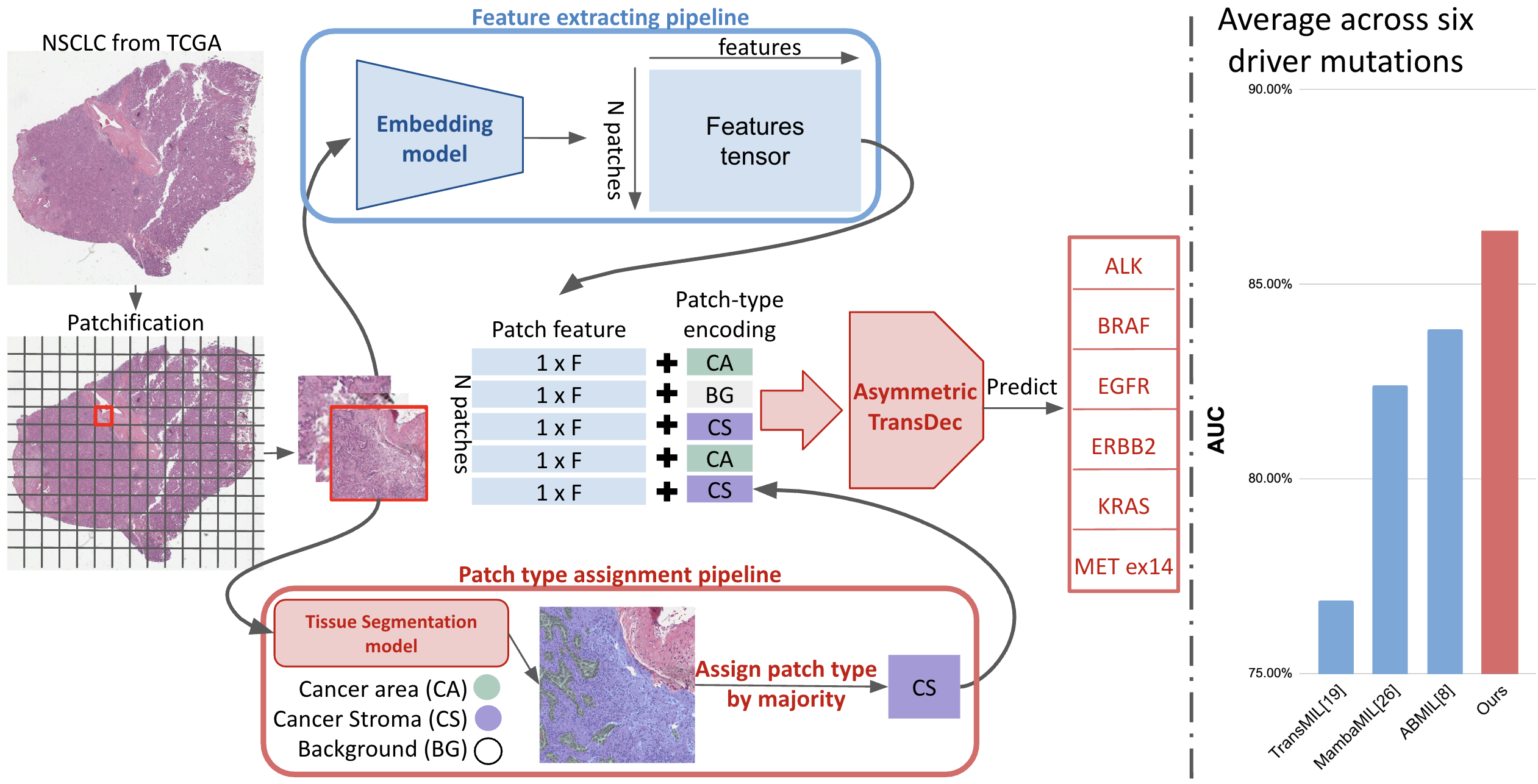}
    \caption{Left) Overview of our MIL approach. First, the WSI is patchified and features for each patch are extracted using an embedding model, typically an FM. Second, each patch is assigned a biologically meaningful tissue type using a segmentation model (stratification). For each tissue type, a learnable encoding is assigned and added to the respective patch. Finally, both features and tissue type are fed into our Asymmetric Transformer Decoder, which is trained to predict a genetic mutation. Right) Average AUROC over predicting 6 mutations for four MIL methods on our internal dataset.
    }
    \label{fig:overview}
\end{figure*}

Advances in genomics along with personalized medicine have transformed the treatment of lung cancer, thereby enhancing the survival rates among patients \cite{herbst2018biology}. The improvement in patient outcomes was significantly influenced by the creation of targeted agents aimed at genetic mutations driving cancer growth \cite{oudkerk2021lung}, such as EGFR and KRAS . These mutations are considered \textit{actionable} as the primary treatment recommendation is dependent on the presence of such mutations. Thus, performing genetic tests for precise identification of actionable mutations is essential for optimal treatment decisions and has been included in international guidelines \cite{planchard2018metastatic}. However, the adoption of next generation sequencing (NGS) testing encounters several obstacles including high costs, the need for specialized infrastructure, long turnaround times, and the necessity for sufficient tissue samples. These issues frequently result in patients receiving suboptimal treatment \cite{habli2020circulating}. In contrast, histopathological analysis of H\&E stained tissue is a standard process in cancer care and readily available within hospitals and other healthcare facilities. Therefore, ML-based tools for the identification of driver mutations in H\&E-stained whole slide images (WSIs) would address some of the limitations of genetic testing. The ML model can function as a screening tool to identify those who are highly unlikely to have a positive driver mutation outcome, thus reducing expenses and unnecessary delays in standard treatment while awaiting a likely negative genetic test \cite{shmatko2022artificial}.

Conventional ML methods are unsuitable for WSIs as their size exceeds GPU memory capacity. Consequently, Multiple Instance Learning (MIL) has emerged as the primary approach for computational pathology (CPath) tasks. Several MIL approaches have been investigated in CPath, including Graph Neural Networks
 \cite{zheng2022graph,shi2024integrative}, attention \cite{ilse2018attention,chen2024benchmarking}, transformers \cite{shao2021transmil,wagner2023transformer}, vision-language 
\cite{shi2024vila}, Mamba \cite{yang2403mambamil}, and others \cite{Zheng_2024_CVPR,WiKG_2024_CVPR}. 
While there are several MIL alternatives, the existing work on the task of lung driver mutation prediction is relatively scarce
\cite{wagner2023transformer}, with the majority focusing on EGFR \cite{campanella2022h,pao2023predicting}.
The scarcity of studies on the automated identification of multiple actionable mutations reveals a notable research gap. In contrast, in this study, we address several clinically relevant lung cancer driver mutations.
 
ML can detect genetic mutations from histology images as driver mutations in cancer correlate strongly with specific histologic phenotypes and alter cell morphology and the tumor micro-environment \cite{correlationMorphology}, visible in tissue samples.
Nevertheless, many MIL studies ignore the semantic significance of each tissue patch and simply use all of them from the WSI. Some approaches select patches with cancer tissue to help the model capture more relevant features. Liu et al.\cite{liu2024semantics} show that focusing the model on particular WSI regions is viable by using tissue segmentation as attention mask labels. 
Our study corroborates the observation that incorporating biologically meaningful information into MIL is powerful, but, in contrast to \cite{liu2024semantics}, we incorporate tissue semantics directly into the model input instead of using it as target.

In summary, we present three contributions. 1) First, a comprehensive benchmark of state-of-the-art CPath MIL methods for predicting clinically relevant NSCLC actionable mutations. 2) Second, a novel integration of tissue biology into transformer-based MIL through tissue-type encoding. 3) Finally, we introduce the Asymmetric Transformer Decoder, a novel architecture that significantly reduces model redundancy, enabling more efficient and accurate mutation prediction.
Fig. \ref{fig:overview} shows an overview of the proposed method.

Crucially, this work aligns with patient and clinical needs for effective targeted therapies within the diagnostic workflow by focusing on the prediction of actionable mutations in NSCLC from widely available H\&E WSIs.

\begin{figure*}[t!]
    \hskip-0.8cm
    \includegraphics[width=1.1\textwidth]{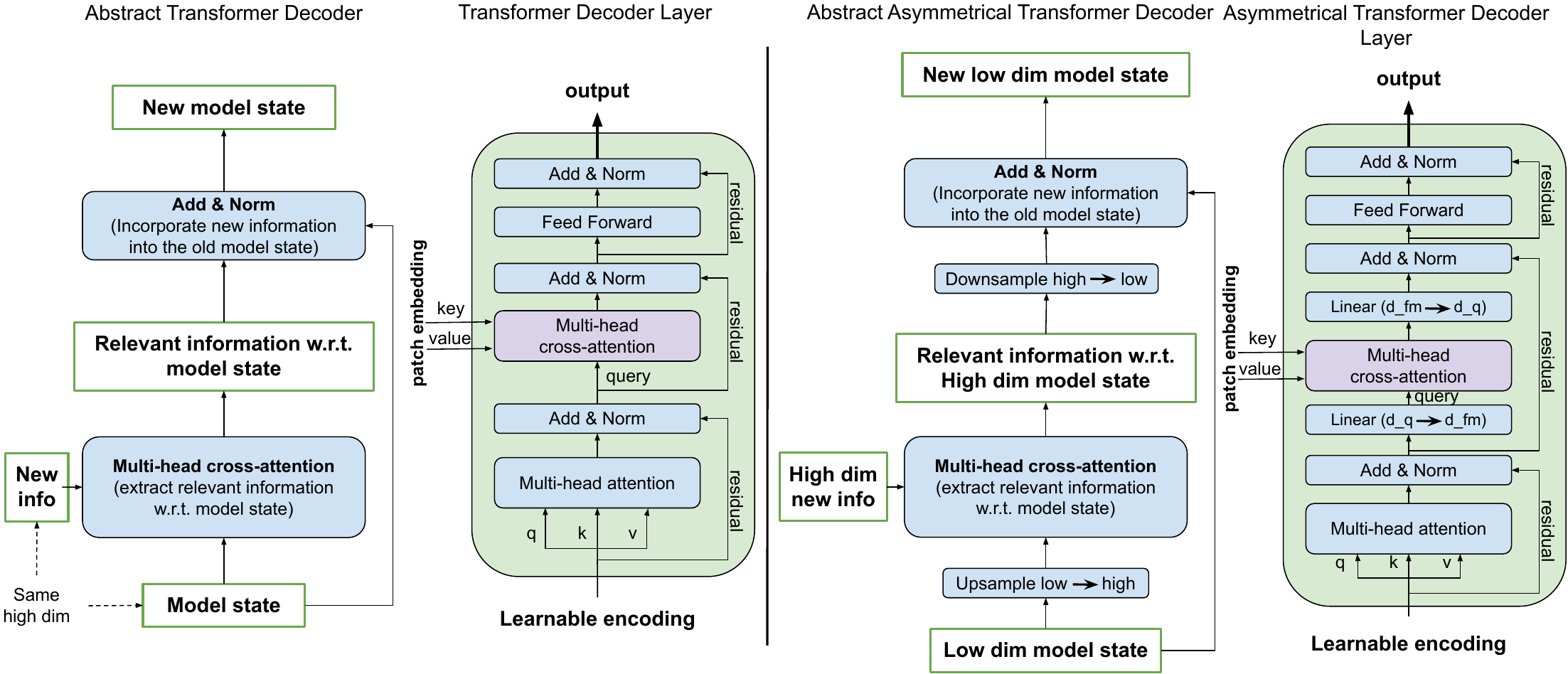}
    \caption{Left) The traditional Transformer Decoder setup for MIL. Right) Our AsymTransDec leverages small learnable parameters in the initial attention module. We then project the query to align with patch embedding dimensions via a linear projection prior to cross attention. The output is reduced back to the small original dimension through a linear mapping before integrating into the residual connection.}
    \label{fig:method}
\end{figure*}

\section{Method}
Bag-wise MIL framework is the standard approach for WSI classification tasks because of the large size of the images. Typically, it involves the following steps: first, each WSI is divided into non-overlapping patches; 
second, a feature extractor is used to map each patch into a lower-dimensional embedding space;
finally, an aggregator combines the patch embeddings and infers a prediction.

\subsubsection{Transformer Decoder for effective information extraction.}
\label{sec:decoder}
Foundation Models (FMs) are trained on extensive data and possess large model sizes, enabling them to produce powerful patch embeddings. The main challenge for an MIL aggregator is to distill pertinent data from a large number of patches into a cohesive representation while reducing overfitting to extraneous details such as source or scanner type that FM may also capture \cite{exaonepath}.
The transformer’s multi-head attention mechanism is ideal for handling long sequences and theoretically well-suited for aggregation. Nevertheless, transformers struggle to meet expectations, still contending with simpler attention models \cite{chen2024benchmarking}.
Despite the fact that FMs already encode patch details, transformer \textbf{encoders} are redundantly used to re-embed representations for downstream tasks \cite{shao2021transmil}, resulting in inefficiency. Conversely, transformer \textbf{decoders} are tailored for pulling out essential features from lengthy encoded vector sequences. 
The transformer decoder uses a set of learned queries to extract data from the encoder output via cross attention without further encoding and transforming the learned patch features, as represented in Fig. \ref{fig:method}-left.
The standard scaled dot-product attention \cite{vaswani2017attention} is given by the following:
\begin{align*}
Attention(Q, K, V) &= \text{softmax}(\frac{QK^T}{\sqrt{d_k}})V
\end{align*}

\noindent where $Q, K, V$ are respectively query, key and value, three learnable encoding vectors of size $d_q$ and $d_{kv}$.
In cross attention, the dimensions of the query, key and value encodings must coincide, i.e. $d_q = d_{kv}$, usually aligning with the high dimensionality of FM embeddings. Increasing the size of query-key-value dimensions can significantly enhance a Transformer model's capacity; however, this may lead to overfitting. Projecting the FM embedding to a smaller dimension is a common approach to mitigate this risk, although it results in notable information loss.
Hence, we opt to maintain full access to the $d_{kv}$ dimension embeddings to prevent such loss. Consequently, an asymmetric approach with different dimensions for query and key-value is necessary.
Under these constraints, we create a model using a compact set of low-dimensional vectors $q$ as queries to lessen overfitting risks. To prevent losing information due to smaller patch embeddings, we transform the query tokens $q$ from a small dimension $d_q$ (e.g., $64$) to a larger dimension $d_{kv}$ (e.g., $1,536$, as in ViT features) for cross attention alignment using a linear projection. The cross attention output is then downscaled to $d_{q}$ and integrated with the query through a residual connection. The detailed architecture is depicted in Fig. \ref{fig:method}-right.
The discrepancy in dimensions between query tokens and patch embeddings (keys, values) is the reason our approach is named Asymmetric Transformer Decoder. The asymmetric cross-attention is

\begin{align*}
AsymmetricAttention(Q_{dq}, K_{dkv}, V_{dkv}) &= \text{softmax}(\frac{Q_{dq}W_{[dq,\ dkv]}K_{dkv}^T}{\sqrt{d_{kv}}})V_{dkv}W_{[dkv,\ dq]}
\end{align*}

\begin{table}[t!]
\footnotesize
\caption{Evaluation of the proposed method with CNN- and ViT-based FMs, and comparison with SotA methods. Results are the average AUROC on 5-fold cross-validation. }
\hskip-1.0cm
\begin{tabular}{|l|l|l|l|l|l|l|l||l|}
\cline{1-9}
Embed & Aggregator & ALK & BRAF & EGFR & ERBB2 & KRAS & METex14 & \textbf{\textit{AVG}} \\ 
\cline{1-9}
 & \#Samples Pos/Neg & 70/1130 & 100/1123 & 275/940 & 78/1145 & 184/1025 & 88/1135 &  \\
\cline{1-9}
CNN & ABMIL\cite{ilse2018attention} & 83.7\% & 80.1\% & 85.1\% & 75.7\% & \textbf{76.5\%} & 80.1\%  & 80.2\% \\
 & TransMIL\cite{shao2021transmil} & 81.2\% & 78.1\% & 83.6\% & 70.9\% & 74.6\% & 77.5\% & 77.6\% \\
 & MambaMIL\cite{yang2403mambamil} & 82.0\% & 80.5\% & 85.5\% & 72.9\% & 74.0\% & 79.8\% & 79.1\% \\
 & Ours & \textbf{85.2\%} & \textbf{82.8\%} & \textbf{86.7\%} & \textbf{78.7\%} & 76.4\% & \textbf{80.8\%} & \textbf{81.8\%} \\
\cline{1-9}
\cline{1-9}
ViT & ABMIL\cite{ilse2018attention} & 87.5\% & 82.4\% & 89.3\% & 75.5\% & 84.2\% & 84.1\% & 83.8\% \\
 & TransMIL\cite{shao2021transmil} & 80.0\% & 79.8\% & 85.8\% & 66.3\% & 73.2\% & 76.2\% & 76.9\% \\
 & MambaMIL\cite{yang2403mambamil} & 88.1\% & 82.2\% & 89.0\% & 73.2\% & 84.2\% & 77.7\% & 82.4\% \\
 & Ours & \textbf{88.4\%} & \textbf{86.5\%} & \textbf{90.8\%} & \textbf{80.1\%} & \textbf{85.9\%} & \textbf{86.4\%} & \textbf{86.3\%} \\
\cline{1-9}
\end{tabular}
\label{tab:benchmark}
\end{table}

\subsubsection{Integrating biological information}
Recent MIL methods have standardized identifying cancer regions within tissue samples \cite{campanella2022h}. Basic image processing is first employed to remove irrelevant patches \cite{campanella2022h}. Subsequently, a segmentation model \cite{unet,deeplabcut} is trained to isolate the cancer area for patch extraction.
This method enriches the MIL input with diagnostically pertinent features; however, we hypothesize that focusing on cancer regions may discard other relevant information. Thus, rather than filtering patches, we suggest providing the model with biological information.
We utilize a DeepLabV3-based \cite{deeplabcut} model to segment three tissue classes: cancer area (CA), cancer stroma (CS), and background (BG). Patches are labeled by the dominant class, defined as occupying over 50\% of the patch. 
A unique learned encoding for each tissue type is added to the respective patch embeddings, as illustrated in Fig. \ref{fig:overview}. We refer to this process as \textbf{tissue-type stratification}.
Importantly, this allows the model to adapt to each patch differently based on their innate biology, without discarding potentially useful information available outside of CA.
Stratification also helps the model to pay attention to each meaningful region, without being biased by the amount of each tissue type within the WSI.
In fact, from each WSI, we randomly sample patches in the ratios of 50\% cancer area, 30\% cancer stroma, and 20\% background.

\begin{table}[t!]
\footnotesize
    \centering
    \caption{Performance of actionable mutation prediction (AUROC) in the TCGA external test set. Models were trained on the complete internal dataset using ViT \cite{hoptimus0}.}
    \begin{tabular}{|l|c|c|c|}
        \hline
        \textbf{Model} & \textbf{EGFR} & \textbf{KRAS} & \textbf{MET} \\ 
        \hline
        ABMIL & 86.2\% & 80.3\% & 83.9\% \\
        TransMIL\cite{shao2021transmil} & 87.3\% & 80.0\% & 72.7\% \\
        MambaMIL\cite{yang2403mambamil} & 86.3\% & 82.2\% & 81.2\% \\
        Ours & \textbf{88.1\%} & \textbf{84.4\%} & \textbf{87.5\%}\\ 
        \hline
    \end{tabular}
    \label{tab:tcga}
\end{table}

\section{Results}

\subsubsection{Dataset}
The dataset comprises 1,223 surgical resection WSIs assessed via NGS. Among all mutations, we utilize six actionable mutations as labels: ALK, BRAF, EGFR, ERBB2, KRAS, and MET\_ex14. Tab. \ref{tab:benchmark} shows the number of available samples for each task.
The WSIs originate from various hospitals
across Western and Asian regions, digitized using Leica Aperio scanners (AT Turbo, AT2, GT450).
An external test set consists of $924$ NSCLC slides from the Cancer Genomic Atlas (TCGA) dataset~\cite{tcga_luad,tcga_lusc}. Due to the scarcity of ALK, BRAF, and ERBB2 positives in TCGA, we focus our testing on EGFR, KRAS, and MET.

\subsubsection{Implementation details and experimental setup}
Existing FMs \cite{kang2022benchmarking,hoptimus0} trained on large unlabeled histopathology datasets served as feature extractors. The models were designed in PyTorch, extracting patch embeddings with 4 NVIDIA T4 GPUs, with size of $224 \times 224$ pixels at $20\times$ magnification.
The embedding dimensions are $1,536$ for ViT and $768$ for CNN. For aggregator training, the AdamW optimizer with a learning rate of $0.0002$ and a cosine annealing scheduler ($T=10$ epochs, $\eta_{\text{min}}=1e-6$) was used. The transformer decoder model utilized a batch size of 128 WSIs. In the transformer model, multi-head attention featured $2$ heads, model dimension (query dimension $d_q$) of $64$, GeLu activation, and dropout ($p=0.5$) before the classifier layer. 

The target mutations are rare; therefore, we develop the method following a stratified 5-fold cross-validation scheme using the internal dataset (1,223 WSIs).  We further mitigate overfitting by tuning the hyper-parameters using only fold 0 from the ALK task and applying them across all tasks and folds. Each mutation is considered a separate task to further tackle class imbalance. In the end, we use the full 1,223 WSIs dataset for training and evaluate in the external test set. We use the area under the receiver operating characteristic curve (AUROC) as the evaluation metric.

\begin{table}[t!]
\footnotesize
\caption{Effect of different tissue types on mutation task performance. Results are the average AUROC on 5-folds. The ViT-based FM \cite{hoptimus0} was used as patch embedding.}
\begin{tabular}{|l|l|l|l|l|l|l||l|}
\cline{1-8}
Tissue type & ALK & BRAF & EGFR & ERBB2 & KRAS & METex14 & \textbf{\textit{AVG}} \\ 
\cline{1-8}
BG-only & 83.7\% & 79.2\% & 84.6\% & 69.4\% & 74.1\% & 82.8\% & 79.0\% \\
CS-only & 87.4\% & 84.9\% & 88.4\% & 77.3\% & 82.7\% & 80.8\% & 83.6\% \\
CA-only & 87.1\% & 84.0\% & 89.2\% & 77.7\% & 83.4\% & 84.7\% & 84.4\% \\
All patches, no strat. & 88.1\% & 85.5\% & 90.4\% & 78.6\% & 85.0\% & 85.5\% & 85.5\% \\
\cline{1-8}
\textbf{All patches, stratified} & \textbf{88.4\%} & \textbf{86.5\%} & \textbf{90.8\%} & \textbf{80.1\%} & \textbf{85.9\%} & \textbf{86.4\%} &  \textbf{86.3\%} \\
\cline{1-8}
\end{tabular}
\label{tab:tissue}
\end{table}

\subsubsection{Detecting actionable driver mutation}

Tab. \ref{tab:benchmark} shows 5-fold cross-validation results across the six actionable mutations. We compare our results with three representative state-of-the-art aggregators: ABMIL \cite{chen2024benchmarking}, TransMIL \cite{shao2021transmil}, and MambaMIL \cite{yang2403mambamil}. We observe that the proposed method outperforms other aggregators in all mutations (except for KRAS with CNN-based FM). Notably, the gap is larger in relation to TransMIL (CNN FM: 4.2\%; ViT FM: 9.4\%), which is also based on Transformer. Indeed, TransMIL obtains the lowest metrics across the considered methods. This suggests that applying Transformers to MIL is not straightforward, requiring a careful formulation, such as the proposed method. We can also observe that TransMIL and MambaMIL achieve better results with the CNN-based FM, while the more lightweight ABMIL can leverage the powerful ViT-based features. In contrast, the proposed method achieves superior results with both FMs. We hypothesize that it is due to the efficient formulation of the proposed Asymmetric Transformer Decoder, which allows it to leverage the higher-dimensional ViT-based features.

We perform external validation of the proposed method on three mutations: EGFR, KRAS, and MET ex14. This is due to the lack of sufficient positive samples in the remaining mutations. Results are presented in Tab. \ref{tab:tcga}. We observe a similar trend as before, where the proposed method obtains the top performance. Moreover, the performance level of our method is consistent, with 88.1\% for EGFR, and 84.4\% for KRAS. In contrast, TransMIL performs well for EGFR prediction (87.3\%), but severely under-performs in MET (72.7\%).

\subsubsection{Ablation study}
Our contribution includes integrating biologically meaningful tissue regions into the model. As shown in Tab. \ref{tab:tissue}, we analyze the contributions of BG, CS, and CA tissues. The CA-only results lead, followed by CS-only and BG-only, in line with expectations. All regions show discriminative capability (see Tab. \ref{tab:benchmark}, BG-only outperforms TransMIL). This might partly result from segmentation errors in miss-assigning patches, and we also propose that cancer, and certain mutations, can alter adjacent tissues. Results without tissue-type stratification, where all patches are equally considered as in standard MIL, outperform considering specific ROI but underperform compared to the proposed tissue-type stratification. This indicates that all WSI regions can aid mutation prediction, yet the stratification method enhances information extraction.

In Tab. \ref{tab:architecture}, we compare the proposed Asymmetric Transformer Decoder aggregator with vanilla Transformer Encoder and Decoder aggregators. The encoder-based aggregator underperforms the decoder-based aggregators by a large margin. The encoder is a large-capacity model aimed at learning powerful representations. However, this leads to difficulties during training and overfitting in a MIL setting with few positive samples. Instead, the decoder is more lightweight and focused on integrating information from the already powerful FM-based features. We observe that even the vanilla decoder obtains better driver mutation prediction performance (84.9\%) than the encoder (78.8\%). The propose Asymmetric Transformer Decoder further boosts the performance to 86.3\%, demonstrating that its efficient design is more adequate as an MIL aggregator.

\begin{table}[t!]
\footnotesize
\caption{Comparison between transformer architectures, using features from the ViT-based FM. TransEnc and TransDec represent the regular transformer encoder and decoder, respectively. Results are the
average AUROC on 5-fold cross-validation.}
\hskip+1.0cm
\begin{tabular}{|l|l|l|l|l|l|l||l|}
\cline{1-8}
Tissue type & ALK & BRAF & EGFR & ERBB2 & KRAS & METex14 & \textbf{\textit{AVG}} \\ 
\cline{1-8}
TransEnc & 86.7\% & 78.8\% & 89.0\% & 66.4\% & 81.1\% & 70.7\% & 78.8\% \\
TransDec & 88.3\% & 82.7\% & 89.9\% & 79.2\% & 84.5\% & 85.0\% & 84.9\% \\
\textbf{AsymTransDec} & \textbf{88.4\%} & \textbf{86.5\%} & \textbf{90.8\%} & \textbf{80.1\%} & \textbf{85.9\%} & \textbf{86.4\%} &  \textbf{86.3\%} \\
\cline{1-8}
\end{tabular}
\label{tab:architecture}
\end{table}

\section{Conclusion}
Identifying driver mutations in lung cancer is essential for targeted therapies and improving patient prognoses. Yet, genetic testing is often slow, expensive, and requires specialized facilities, limiting its use and delaying treatment in many clinics. To overcome these barriers, we developed the Asymmetric Transformer Decoder model, an MIL-based technique that examines typical H\&E-stained whole-slide images (WSIs) to identify mutations. Unlike conventional methods focusing only on cancer areas, our model includes tissue-type embeddings to further utilize data from cancer stroma and background regions, boosting mutation detection. Experiments reveal that our model surpasses leading MIL models, notably for rare mutations like ERBB2 and BRAF. This indicates its potential as a scalable, cost-effective genetic testing alternative suitable for local deployment in various clinical contexts, enhancing access to precision oncology. 
\newpage

\bibliography{main}

\begin{thebibliography}{27}
\providecommand{\natexlab}[1]{#1}
\providecommand{\url}[1]{\texttt{#1}}
\expandafter\ifx\csname urlstyle\endcsname\relax
  \providecommand{\doi}[1]{doi: #1}\else
  \providecommand{\doi}{doi: \begingroup \urlstyle{rm}\Url}\fi

\bibitem[Albertina et~al.(2016)Albertina, Watson, Holback, Jarosz, Kirk, Lee, Rieger-Christ, and Lemmerman]{tcga_luad}
B.~Albertina, M.~Watson, C.~Holback, R.~Jarosz, S.~Kirk, Y.~Lee, K.~Rieger-Christ, and J~Lemmerman.
\newblock The cancer genome atlas lung adenocarcinoma collection (tcga-luad) (version 4) [data set].
\newblock The Cancer Imaging Archive. doi: 10.7937/K9/TCIA.2016.JGNIHEP5, 2016.

\bibitem[Campanella et~al.(2022)Campanella, Ho, H{\"a}ggstr{\"o}m, Becker, Chang, Vanderbilt, and Fuchs]{campanella2022h}
Gabriele Campanella, David Ho, Ida H{\"a}ggstr{\"o}m, Anton~S Becker, Jason Chang, Chad Vanderbilt, and Thomas~J Fuchs.
\newblock H\&e-based computational biomarker enables universal egfr screening for lung adenocarcinoma.
\newblock \emph{arXiv preprint arXiv:2206.10573}, 2022.

\bibitem[Chen et~al.(2024)Chen, Campanella, Elmas, Stock, Zeng, Polydorides, Schoenfeld, Huang, Houldsworth, Vanderbilt, and Fuchs]{chen2024benchmarking}
Shengjia Chen, Gabriele Campanella, Abdulkadir Elmas, Aryeh Stock, Jennifer Zeng, Alexandros~D. Polydorides, Adam~J. Schoenfeld, Kuan-lin Huang, Jane Houldsworth, Chad Vanderbilt, and Thomas~J. Fuchs.
\newblock Benchmarking embedding aggregation methods in computational pathology: A clinical data perspective.
\newblock \emph{Proceedings of the MICCAI Workshop on Computational Pathology}, 254, 2024.

\bibitem[Habli et~al.(2020)Habli, AlChamaa, Saab, Kadara, and Khraiche]{habli2020circulating}
Zeina Habli, Walid AlChamaa, Raya Saab, Humam Kadara, and Massoud~L Khraiche.
\newblock Circulating tumor cell detection technologies and clinical utility: Challenges and opportunities.
\newblock \emph{Cancers}, 12\penalty0 (7), 2020.

\bibitem[Herbst et~al.(2018)Herbst, Morgensztern, and Boshoff]{herbst2018biology}
Roy~S Herbst, Daniel Morgensztern, and Chris Boshoff.
\newblock The biology and management of non-small cell lung cancer.
\newblock \emph{Nature}, 553\penalty0 (7689):\penalty0 446--454, 2018.

\bibitem[Ilse et~al.(2018)Ilse, Tomczak, and Welling]{ilse2018attention}
Maximilian Ilse, Jakub Tomczak, and Max Welling.
\newblock Attention-based deep multiple instance learning.
\newblock \emph{International conference on machine learning}, 2018.

\bibitem[Kang et~al.(2023)Kang, Song, Park, Yoo, and Pereira]{kang2022benchmarking}
Mingu Kang, Heon Song, Seonwook Park, Donggeun Yoo, and Sérgio Pereira.
\newblock Benchmarking self-supervised learning on diverse pathology datasets.
\newblock \emph{Proceedings of the IEEE/CVF Conference on Computer Vision and Pattern Recognition (CVPR)}, June 2023.

\bibitem[Kirk et~al.(2016)Kirk, Lee, Kumar, Filippini, Albertina, Watson, Rieger-Christ, and Lemmerman]{tcga_lusc}
S.~Kirk, Y.~Lee, P.~Kumar, J.~Filippini, B.~Albertina, M.~Watson, K.~Rieger-Christ, and J~Lemmerman.
\newblock The cancer genome atlas lung squamous cell carcinoma collection (tcga-lusc) (version 4) [data set].
\newblock The Cancer Imaging Archive. doi: 10.7937/K9/TCIA.2016.TYGKKFMQ, 2016.

\bibitem[Li et~al.(2024)Li, Chen, Chu, Sun, Guan, Han, and He]{WiKG_2024_CVPR}
Jiawen Li, Yuxuan Chen, Hongbo Chu, Qiehe Sun, Tian Guan, Anjia Han, and Yonghong He.
\newblock Dynamic graph representation with knowledge-aware attention for histopathology whole slide image analysis.
\newblock \emph{Proceedings of the IEEE/CVF Conference on Computer Vision and Pattern Recognition (CVPR)}, June 2024.

\bibitem[Liu et~al.(2024)Liu, Wu, Elmore, and Shapiro]{liu2024semantics}
Kechun Liu, Wenjun Wu, Joann~G Elmore, and Linda~G Shapiro.
\newblock Semantics-aware attention guidance for diagnosing whole slide images.
\newblock \emph{International Conference on Medical Image Computing and Computer-Assisted Intervention (MICCAI)}, 2024.

\bibitem[Mathis et~al.(2018)Mathis, Mamidanna, Cury, Abe, Murthy, Mathis, and Bethge]{deeplabcut}
Alexander Mathis, Pranav Mamidanna, Kevin~M. Cury, Taiga Abe, Venkatesh~N. Murthy, Mackenzie~W. Mathis, and Matthias Bethge.
\newblock Deeplabcut: markerless pose estimation of user-defined body parts with deep learning.
\newblock \emph{Nature Neuroscience}, 2018.

\bibitem[Oudkerk et~al.(2021)Oudkerk, Liu, Heuvelmans, Walter, and Field]{oudkerk2021lung}
Matthijs Oudkerk, ShiYuan Liu, Marjolein~A Heuvelmans, Joan~E Walter, and John~K Field.
\newblock Lung cancer ldct screening and mortality reduction—evidence, pitfalls and future perspectives.
\newblock \emph{Nature reviews Clinical oncology}, 18\penalty0 (3):\penalty0 135--151, 2021.

\bibitem[Pao et~al.(2023)Pao, Biggs, Duncan, Lin, Davis, Huang, Ferguson, Janovitz, Hiemenz, Eddy, et~al.]{pao2023predicting}
James~J Pao, Mikayla Biggs, Daniel Duncan, Douglas~I Lin, Richard Davis, Richard~SP Huang, Donna Ferguson, Tyler Janovitz, Matthew~C Hiemenz, Nathanial~R Eddy, et~al.
\newblock Predicting egfr mutational status from pathology images using a real-world dataset.
\newblock \emph{Scientific reports}, 13\penalty0 (1), 2023.

\bibitem[Planchard et~al.(2018)Planchard, Popat, Kerr, Novello, Smit, Faivre-Finn, Mok, Reck, Van~Schil, Hellmann, et~al.]{planchard2018metastatic}
D~Planchard, ST~Popat, K~Kerr, S~Novello, EF~Smit, Corinne Faivre-Finn, TS~Mok, M~Reck, PE~Van~Schil, MD~Hellmann, et~al.
\newblock Metastatic non-small cell lung cancer: Esmo clinical practice guidelines for diagnosis, treatment and follow-up.
\newblock \emph{Annals of Oncology}, 29, 2018.

\bibitem[Ronneberger et~al.(2015)Ronneberger, Fischer, and Brox]{unet}
Olaf Ronneberger, Philipp Fischer, and Thomas Brox.
\newblock U-net: Convolutional networks for biomedical image segmentation.
\newblock \emph{International Conference on Medical Image Computing and Computer-Assisted Intervention (MICCAI)}, pages 234--241, 2015.

\bibitem[Saillard et~al.(2024)Saillard, Jenatton, Llinares-López, Mariet, Cahané, Durand, and Vert]{hoptimus0}
Charlie Saillard, Rodolphe Jenatton, Felipe Llinares-López, Zelda Mariet, David Cahané, Eric Durand, and Jean-Philippe Vert.
\newblock H-optimus-0, 2024.
\newblock URL \url{https://github.com/bioptimus/releases/tree/main/models/h-optimus/v0}.

\bibitem[Shao et~al.(2021)Shao, Bian, Chen, Wang, Zhang, Ji, et~al.]{shao2021transmil}
Zhuchen Shao, Hao Bian, Yang Chen, Yifeng Wang, Jian Zhang, Xiangyang Ji, et~al.
\newblock Transmil: Transformer based correlated multiple instance learning for whole slide image classification.
\newblock \emph{Advances in neural information processing systems}, 34, 2021.

\bibitem[Shi et~al.(2024{\natexlab{a}})Shi, Li, Gong, Zheng, and Fu]{shi2024vila}
Jiangbo Shi, Chen Li, Tieliang Gong, Yefeng Zheng, and Huazhu Fu.
\newblock Vila-mil: Dual-scale vision-language multiple instance learning for whole slide image classification.
\newblock \emph{Proceedings of the IEEE/CVF Conference on Computer Vision and Pattern Recognition}, 2024{\natexlab{a}}.

\bibitem[Shi et~al.(2024{\natexlab{b}})Shi, Zhang, Kong, and Wang]{shi2024integrative}
Zhan Shi, Jingwei Zhang, Jun Kong, and Fusheng Wang.
\newblock Integrative graph-transformer framework for histopathology whole slide image representation and classification.
\newblock \emph{International Conference on Medical Image Computing and Computer-Assisted Intervention (MICCAI)}, 2024{\natexlab{b}}.

\bibitem[Shmatko et~al.(2022)Shmatko, Ghaffari~Laleh, Gerstung, and Kather]{shmatko2022artificial}
Artem Shmatko, Narmin Ghaffari~Laleh, Moritz Gerstung, and Jakob~Nikolas Kather.
\newblock Artificial intelligence in histopathology: enhancing cancer research and clinical oncology.
\newblock \emph{Nature cancer}, 3\penalty0 (9), 2022.

\bibitem[Vaswani et~al.(2017)Vaswani, Shazeer, Parmar, Uszkoreit, Jones, Gomez, Kaiser, and Polosukhin]{vaswani2017attention}
Ashish Vaswani, Noam Shazeer, Niki Parmar, Jakob Uszkoreit, Llion Jones, Aidan~N Gomez, {\L}ukasz Kaiser, and Illia Polosukhin.
\newblock Attention is all you need.
\newblock \emph{Advances in neural information processing systems}, 30, 2017.

\bibitem[Villa et~al.(2014)Villa, Cagle, Johnson, Patel, Yeldandi, Raj, DeCamp, and Raparia]{correlationMorphology}
Celina Villa, Philip~T Cagle, Melissa Johnson, Jyoti~D Patel, Anjana~V Yeldandi, Rishi Raj, Malcolm~M DeCamp, and Kirtee Raparia.
\newblock Correlation of egfr mutation status with predominant histologic subtype of adenocarcinoma according to the new lung adenocarcinoma classification of the international association for the study of lung cancer/american thoracic society/european respiratory society.
\newblock \emph{Archives of Pathology and Laboratory Medicine}, 138\penalty0 (10), 2014.

\bibitem[Wagner et~al.(2023)Wagner, Reisenb{\"u}chler, West, Niehues, Zhu, Foersch, Veldhuizen, Quirke, Grabsch, van~den Brandt, et~al.]{wagner2023transformer}
Sophia~J Wagner, Daniel Reisenb{\"u}chler, Nicholas~P West, Jan~Moritz Niehues, Jiefu Zhu, Sebastian Foersch, Gregory~Patrick Veldhuizen, Philip Quirke, Heike~I Grabsch, Piet~A van~den Brandt, et~al.
\newblock Transformer-based biomarker prediction from colorectal cancer histology: A large-scale multicentric study.
\newblock \emph{Cancer Cell}, 41\penalty0 (9), 2023.

\bibitem[Yang et~al.(2024)Yang, Wang, and Chen]{yang2403mambamil}
Shu Yang, Yihui Wang, and Hao Chen.
\newblock Mambamil: Enhancing long sequence modeling with sequence reordering in computational pathology.
\newblock \emph{International Conference on Medical Image Computing and Computer-Assisted Intervention (MICCAI)}, 2024.

\bibitem[Yun et~al.(2024)Yun, Hu, Kim, Jang, and Lee]{exaonepath}
Juseung Yun, Yi~Hu, Jinhyung Kim, Jongseong Jang, and Soonyoung Lee.
\newblock Exaonepath 1.0 patch-level foundation model for pathology.
\newblock \emph{arXiv preprint arXiv:2408.00380}, 2024.

\bibitem[Zheng et~al.(2024)Zheng, Jiang, and Yao]{Zheng_2024_CVPR}
Tingting Zheng, Kui Jiang, and Hongxun Yao.
\newblock Dynamic policy-driven adaptive multi-instance learning for whole slide image classification.
\newblock \emph{Proceedings of the IEEE/CVF Conference on Computer Vision and Pattern Recognition (CVPR)}, June 2024.

\bibitem[Zheng et~al.(2022)Zheng, Gindra, Green, Burks, Betke, Beane, and Kolachalama]{zheng2022graph}
Yi~Zheng, Rushin~H Gindra, Emily~J Green, Eric~J Burks, Margrit Betke, Jennifer~E Beane, and Vijaya~B Kolachalama.
\newblock A graph-transformer for whole slide image classification.
\newblock \emph{IEEE transactions on medical imaging}, 41\penalty0 (11), 2022.

\end{thebibliography}

\end{document}